\documentstyle [12pt]{article}
\newcommand{\nwc}{\newcommand}
\nwc{\be}  {\begin{equation}}
\nwc{\ee}  {\end{equation}}
\nwc{\ba}  {\begin{array}}
\nwc{\ea}  {\end{array}}
\nwc{\bdm} {\begin{displaymath}}
\nwc{\edm} {\end{displaymath}}
\nwc{\bea} {\be\ba{rcl}}
\nwc{\eea} {\ea\ee}
\nwc{\ben} {\begin{eqnarray}}
\nwc{\een} {\end{eqnarray}}
\newcommand{\eins}{  1\!{\rm l}  }
\def\KK{{\rm I\kern -.2em  K}}
\def\NN{{\rm I\kern -.16em N}}
\def\RR{{\rm I\kern -.2em  R}}
\def\ZZ{Z \kern -.43em Z}
\def\QQ{{\rm \kern .25em
             \vrule height1.4ex depth-.12ex width.06em\kern-.31em Q}}
\def\CC{{\rm \kern .25em
             \vrule height1.4ex depth-.12ex width.06em\kern-.31em C}}
\def\ZZZ{Z\kern -0.31em Z}

\begin{document}
\begin{titlepage}
\begin{flushright}
HD--THEP--98--61
\end{flushright}
\quad\\
\vspace{1.8cm}

\begin{center}
{\Large Natural Maximal $\nu_\mu-\nu_\tau$ Mixing}\\
\vspace{2cm}
Christof Wetterich\footnote{e-mail: C.Wetterich@thphys.uni-heidelberg.de}\\
\bigskip
Institut  f\"ur Theoretische Physik\\
Universit\"at Heidelberg\\
Philosophenweg 16, D-69120 Heidelberg\\
\vspace{3cm}

\end{center}

\begin{abstract}
The naturalness of maximal mixing between myon- and tau-neutri\-nos
is investigated. A spontaneously broken nonabelian generation
symmetry can explain a small parameter which governs the deviation
from maximal mixing. In many cases all three
neutrino masses are almost degenerate. Maximal $\nu_\mu-\nu_\tau$-mixing
would indicate that the leading contribution to the light neutrino
masses arises from the expectation value of a heavy weak triplet
rather than from the seesaw mechanism. In this scenario the
deviation from maximal mixing is predicted to be less than about
1\%.
\end{abstract}
\end{titlepage}

\newpage
\vspace{1cm}

Recent measurements of the $\nu_e$ and $\nu_\mu$ yields of atmospheric
neutrinos \cite{1} give a strong hint for $\nu_\mu-\nu_\tau$ oscillations
characterized by a difference in squared mass $\Delta m^2_a\approx
5\cdot 10^{-3} eV^2$ and large mixing angle $\sin^22\vartheta \stackrel
{\scriptstyle>}{\sim} 0.8$. In comparison with the charged fermions
the large generation mixing may surprise at first sight. In fact, it
can arise quite naturally if the pattern of neutrino masses and
mixings is connected to some spontaneously broken generation
symmetry. A first investigation of abelian
generation symmetries which could explain the hierarchies
in the charged fermion masses has revealed \cite{Bijnens}
that often the generation charge
of the myon- and tau-neutrino is the same, $Q(\nu_\mu)=Q(\nu_\tau)$.
In consequence the mixing between $\nu_\mu$ and $\nu_\tau$ is
not suppressed by any small parameter in contrast to the mixing in
the charged fermion sector. For such models large
$\nu_\mu-\nu_\tau$-mixing has been predicted \cite{Bijnens},\cite{Ram}.
In this approach, all small mass ratios and mixing angles are explained
in terms of various powers of a small parameter connected to generation
symmetry breaking. All realistic
models with generation group $U(1)$ and unification group
$SU(5)$ were found to have $Q(\nu_\mu)=Q(\nu_\tau)$.
Furthermore, this type of models has recently been investigated in the
context of cosmological leptogenesis \cite{Yanagida}. Some of
the models lead to a cosmological baryon asymmetry in the observed
order of magnitude.

Typically, in models with an abelian generation
group $\sin^22\vartheta$ comes out
naturally of order one, but not necessarily very close to one.
In this letter we ask: Are there natural models where $\sin^2
2\vartheta=1-\epsilon$ with $\epsilon\ll 1$? Such a ``maximal
$\nu_\mu-\nu_\tau$-mixing'' is compatible with experiment, but a
really small parameter, say $\epsilon<0.1$, is not established.
We want to show that maximal $\nu_\mu-\nu_\tau$-mixing, if found
in nature, would imply very interesting consequences for our
understanding of possible generation symmetries and their breaking
pattern. In particular, generation symmetries explaining naturally
$\epsilon\ll 1$ must be nonabelian.

For the left-handed neutrinos of the standard model the smallness
of their masses as compared to the charged fermion masses is naturally
explained \cite{Magg} as a consequence of the gauge hierarchy,
i.e. the small ratio between the Fermi scale $\sim M_W$ and the
unification scale $\sim M$. Since the only allowed mass term is
a symmetric neutrino bilinear, it must transform as a triplet under the
weak gauge group $SU(2)_L$. No renormalizable operator of this type
exists in the standard model (and many extensions of it). Allowed
nonrenormalizable operators $\sim \frac{1}{M}\nu\nu dd$ involve the
vacuum expectation value $d$ of the Higgs doublet twice. They are
suppressed, however, by the inverse of some large scale $M$, explaining
naturally why $m_\nu\sim M^2_W/M$ is much smaller than a charged fermion
mass which is $\sim d\sim M_W$ \cite{Magg, Witten, Shafi, W1}. In
many models $M$ is the scale of spontaneous $B-L$ symmetry breaking.
The seesaw-mechanism \cite{seesaw} can be understood
naturally in this context if the dominant contribution to the
nonrenormalizable operator comes from the exchange of superheavy
neutrinos\footnote{We use here a notation where all particles
are named as left-handed particles. For example, the left-handed
antineutrino $\nu^c$ is conjugate to the right-handed neutrino.} $\nu^c$.
In this case $M$ is associated with the heavy neutrino masses.
The seesaw mechanism is, however, only a partial ingredient for an
explanation of the smallness of the neutrino masses. We will actually
see below that it does not give the dominant contribution to the neutrino
masses in case of maximal $\nu_\mu-\nu_\tau$-mixing. The dominant
mass term arises instead from the ``induced
triplet mechanism'', i.e. the expectation value of a heavy
scalar $SU(2)_L$-triplet \cite{Magg} which is naturally
suppressed\footnote{See ref. \cite{MS} for an application of the
induced triplet mechanism
\cite{Magg} for a detailed discussion of triplet potentials in
left-right symmetric models. A general discussion of mixing matrices
in presence of triplet expectation values can be found in \cite{triplet}.}
$\sim M_W^2/M$.

We restrict
the discussion here to models where the smallness of the neutrino
masses is directly related to the gauge hierarchy. In particular, we
assume that all fermions which are not protected by the chiral
$SU(2)_L\times U(1)_Y$ gauge symmetry are superheavy and we do
not consider generation symmetries which are only broken at the Fermi
scale. This implies that there are only three light neutrinos
$\nu_e, \nu_\mu,\nu_\tau$ without any ``sterile neutrinos''.

The orders of magnitude $m_f\sim M_W$,
$m_\nu\sim M^2_W/M$ for the charged and neutral fermions,
respectively, reflect only the very rough structure which is
induced by the small scale of $SU(2)_L$-symmetry breaking.
We next pursue the idea that all small quantities in the mass
matrices are dictated by symmetry in order to understand the
generation fine structure responsible for the substantial splitting
within the charged or neutral fermion masses. A generation symmetry
can differentiate between the electron and the $\tau$
or the up and the top quark since those may transform differently.
The spontaneous breaking of the generation group induces a small
parameter $\lambda$ (or several such parameters) \cite{Frogatt},
\cite{Dimopoulos}, \cite{W2}, \cite{B1}.
It corresponds to the ratio $M_G/M$ with
$M_G$ a characteristic scale for the spontaneous breaking of
the generation group. We deal here with a fine structure around the
unification scale with typical orders of magnitude $\lambda=M_G/M
\approx 10^{-2}-10^{-1}$. Depending on the generation charges
of the various fermion bilinears which appear in the fermion
mass matrices a certain number $p$ of generation symmetry breaking
operators is needed in order to construct a singlet. In consequence,
the corresponding matrix element of a charged fermion mass matrix
is proportional $\sim \lambda^pM_W$. The phenomenologically
required powers of $\lambda$
have been discussed systematically in \cite{B2}. We may call the doublet
which is responsible for the spontaneous symmetry breaking of
$SU(2)_L$ in the limit of unbroken generation symmetry
the ``leading doublet''. Typically, it only contributes to the top quark
mass \cite{W2},\cite{B1}.
Due to generation symmetry breaking, the low mass doublet in the
sense of an effective field theory acquires an admixture of other
doublets $\sim \lambda^p$.

These ideas were used for an understanding of the mass matrix
$M_\nu$ for the light neutrinos \cite{Bijnens}. For an abelian
generation symmetry consistent with the structure of the charged
fermion mass matrices interesting patterns for $M_\nu$ were found
and two general lessons became visible: (i) Typically, the
mass and mixing pattern for the neutrinos is not similar to the
one for the charged fermions. The reason is that generically
the $SU(2)_L$-triplet operators entering $M_\nu$ transform
differently under the generation group as compared to the
$SU(2)_L$-doublet operators responsible for the charged
fermion masses. (ii) The proportionality of the neutrino masses
to the squares of the charged fermion masses of the same generation,
i.e. $m_{\nu_i}/m_{\nu_j}=m^2_{f_i}/m^2_{f_j}$ is usually not
realized.

The discussion in ref. \cite{Bijnens} was
restricted to an abelian generation symmetry. In this context
a large $\nu_\mu-\nu_\tau$-mixing is often natural, but maximal
mixing in the sense of $\epsilon\ll 1$
would be difficult to understand. As an alternative
we will see in this note that a non-abelian generation symmetry can
indeed explain naturally a neutrino mass pattern where

\begin{description}
\item[(i)] the $\nu_\mu-\nu_\tau$-mixing is maximal, and

\item[(ii)] the mass-squared difference between $\nu_e$ and $\nu_A
=\frac{1}{\sqrt 2}(\nu_\mu+\nu_\tau)$ is much smaller than
the mass-squared difference with the third neutrino $\nu_B
=-\frac{i}{\sqrt2}(\nu_\mu-\nu_\tau)$.
\end{description}
Here the pattern of mass eigenstates may either be hierarchical  
$|m_{\nu_e}|\ll|m_{\nu_A}|\ll|m_{\nu_B}|$ or degenerate.
In the degenerate case all three masses are either almost equal
and larger than $(\Delta m^2_a)^{1/2}$ or they are typically all
of the order $(\Delta m^2_a)^{1/2}$. The simplest generation
symmetry leading to such a pattern can be generated from the
discrete reflections $R:\ \nu_e
\leftrightarrow \nu_A$ and $T:\ (\nu_\mu\to -\nu_\mu,\ \nu_\tau
\to -\nu_\tau)$.

It is instructive to look first at the
muon- and tau-neutrinos neglecting
all other particles. The most general (Majorana) mass matrix
is symmetric, and for a first approach we also take it to be real
\be\label{1}
M_{\nu,2}=\left(\begin{array}{cc}
a-c,& b\\
b\ ,&a+c\end{array}\right)\ee
Almost maximal mixing requires a small ratio $|c/b|\ll1$ with
\be\label{2}
\sin^2(2\vartheta)=1-\frac{c^2}{b^2}+0\left(\frac{c^4}{b^4}\right)\ee
It is not difficult to conceive a mechanism leading
naturally to $|c/b|\ll1$. An example is a generation symmetry
that forbids the diagonal elements in $M_{\nu,2}$. They can then be induced
only by spontaneous breaking of such a symmetry, leading to
a suppression factor $c/b\sim(M_G/M)^p$. Any symmetry under which
$\nu_\mu$ and $\nu_\tau$ transform differently while the
complex conjugate of the leading $SU(2)_L$-triplet operator shares the
same generation quantum numbers as the bilinear $\nu_\mu\nu_\tau$ will
be sufficient for this purpose.

Further restrictions arise if we take into account
information about the electron neutrino from solar neutrino observations.
An MSW explanation \cite{MSW} of the solar neutrino puzzle by neutrino oscillations
requires that the relevant mass-squared difference
$\Delta m^2_s\approx 5\cdot 10^{-6}eV^2$ is small as compared
to $\Delta m^2_a\approx 5\cdot 10^{-3}eV^2$. For three neutrinos
this leaves only two possibilities for the pattern of neutrino mass
eigenvalues, namely
\[{\rm (A)}\qquad |m_1|\ \ll\ |m_2|\approx \sqrt{\Delta m^2_s}
\ \ll\ |m_3|\approx \sqrt{\Delta m^2_a}\]
or
\[{\rm (B)}\qquad |m_1|\approx |m_2|\approx |m_3|\approx m\quad,\quad
\Delta m^2_s\ll \Delta m^2_a <m^2\]
We concentrate on the case of small mixing of the electron neutrino
with the other neutrinos. The mass matrix $M_{\nu,2}$ (eq.(\ref{1})) is
then a reasonable approximation for the $\nu_\mu-\nu_\tau$
sector. We denote the eigenvalue with the larger (smaller)
absolute size by $m_+(m_-)$ and introduce the ratio
\be\label{3}
R=\frac{m^2_+-m^2_-}{m^2_++m^2_-}=
\frac{2|a|\sqrt{b^2+c^2}}{a^2+b^2+c^2}
\approx\frac{2|ab|}{a^2+b^2}\ee
The ``hierarchical mass pattern'' (A) needs $R$ very close
to one or
\be\label{4}
a=\pm b(1+\eta)\ee
In this case we need a symmetry explanation for a very small
difference $|a|-|b|$, i.e. $|\eta|=2
\sqrt{\Delta m^2_s/\Delta m^2_a}$.
Whenever $|\eta|$ exceeds this value, the only alternative
is the ``degenerate mass pattern'' (B). A leading form of the
neutrino mass matrix with maximal mixing is
\be\label{5}
M_{max}=\left(\begin{array}{ccc}
\pm a\pm b&0&0\\
0&a&b\\
0&b&a\end{array}\right)\ee
It has two degenerate mass eigenvalues $|m|=|a\pm b|$ and one eigenvalue
$|m|=|a\mp b|$. For
$(M_{max})_{11}=0\ (|a|=|b|)$ one has the hierarchical pattern (A).
On the other hand, for $|a|\ll|b|$ one finds three almost
degenerate values for the neutrino masses\footnote{Degenerate
neutrino masses without connection to maximal $\nu_\mu-\nu_\tau$-mixing
have been discussed earlier \cite{DEGEN}, often in a setting
where $|a|\gg|b|$.}
which are substantially
larger than $\sqrt{\Delta m^2_a}$
\be\label{6}
\frac{\Delta m^2_a}{m^2}\approx 4\left|\frac{a}{b}\right|\ee
This cosmologically interesting scenario extends
qualitatively to the case where  $a$ and $c$ are of similar
magnitude. Adding the pieces $\sim c$ in the $\nu_\mu-
\nu_\tau$ sector induces a contribution $\Delta m^2_s\approx c^2$.
For $c=a$ this yields
$m^2/\Delta m^2_a=\frac{1}{16}\Delta m^2_a/\Delta m_s^2$.

These considerations can be generalized for complex symmetric
matrices. For a $\nu_\mu-\nu_\tau$-matrix  $M_{\nu,2}$ (1) the
condition for maximal mixing is now $|Re(ac^*)|
\ \ll \ [(Re(b^*a))^2+(Im(b^*c))^2]^{1/2}=0$.
As an example, a leading mass matrix
\be\label{A}
M_A=m Y,\quad Y=\frac{1}{2}\left(\begin{array}{ccc}
0&0&0\\ 0&1&i\\ 0&i&-1\end{array} \right)\ee
leads to maximal mixing with a hierarchical mass pattern
(A) with eigenvalues $(0,0,m)$.
A matrix of the type (5) with two degenerate eigenvalues and maximal
mixing can be written\footnote{Due to the freedom in the choice
of phases the matrices $M_A$ and $M_{max}$ are not the most
general ones for maximal mixing and given mass eigenvalues.} as
\be\label{A1}
M_{max}=bW+a\eins,\quad W=\left(
\begin{array}{ccc}1&0&0\\ 0&0&1\\ 0&1&0\end{array}\right)\ee
with complex $b$ and $a$. Here the mass difference obeys
$\Delta m^2_a=4|Re(ba^*)|$ and for $|Re(ba^*)\ |\ll\
|b|^2$ all three masses are approximately degenerate.

Let us next investigate possible generation symmetries which
lead to the degenerate mass pattern $M_B=mW$.
A global $U(1)$ symmetry with charges $(q_e,q_\mu, q_\tau)$ for
$(\nu_e, \nu_\mu, \nu_\tau)$ where $2q_e=q_\mu+q_\tau$ and
$q_\mu\not= q_\tau$ enforces a mass matrix
\be\label{8}
\hat M_B=\left(\begin{array}{ccc}
m_1&0&0\\ 0& 0& m\\ 0&m&0\end{array}\right)\ee
provided the generation charge of the leading $SU(2)_L$ triplet is
$-2q_e$. This generalizes to discrete subgroups as, for example,
the $Z_3$-symmetry $\nu_e\to e^{i\varphi}\nu_e,\ \nu_\mu\to
e^{2i\varphi}\nu_\mu,\ \nu_\tau\to\nu_\tau,\ t\to
e^{-2i\varphi}t,\ \varphi=\frac{2\pi}{3}$. Symmetries
enforcing $m_1=m$ can be found most easily by a unitary change of basis
$\nu=U\tilde\nu$, i.e.
\be\label{9}
(\nu_e,\nu_\mu,\nu_\tau)=(\nu_e, \nu_A, \nu_B)U^T, \quad
U=\frac{1}{\sqrt2}\left(\begin{array}{ccc}
\sqrt2&0&0\\ 0&1&-i\\ 0&1& i\end{array}\right),\ee
which makes the neutrino mass matrix diagonal,
$U^T\hat M_BU=diag(m_1,m,m)$.
For $m_1=m$ one has $U^*V^TU^TM_BUVU^\dagger=M_B$ if $V$ is an element
of $O(3), V^TV=1$. Consider now transformations $V$ which form
a subgroup of $O(3)$ such that the unit matrix is the only
invariant symmetric $3\times 3$ matrix. Then the generation
symmetry $\nu\to S\nu,\ S=UVU^\dagger$,
enforces the mass matrix $M_B=mW$ provided the leading $SU(2)_L$
triplet transforms trivially.

A possible (minimal) discrete transformation group can be
constructed from the elements
\be\label{10}
R=\frac{1}{2}\left(\begin{array}{ccc}
0&\sqrt2&\sqrt2\\ \sqrt2&1&-1\\ \sqrt2&-1& 1\end{array}\right),
\quad T(\varphi)=\left(\begin{array}{ccc}
1&0&0\\ 0&e^{i\varphi}&0\\ 0&0& e^{-i\varphi}\end{array}\right)
\ee
with $\varphi_n=2\pi/n,\ n\in{\NN},\ n\geq2$. The combination of $R$ and
$T_n\equiv T(\varphi_n)$ enforces a matrix of the form
$M_{max}$ (eq. (\ref{A1})) if the triplet transforms trivially.
For $n>2$ it implies
furthermore $a=0$. We note that $R$ corresponds to $\nu_e\leftrightarrow
\nu_A=\frac{1}{\sqrt2}(\nu_\mu+\nu_\tau)$ with $R^2=1$. Both
$R$ and $T(\varphi)$ are elements of $O(3)$ obeying the relation
$S^T=(UU^T)^*S^{-1}UU^T=WS^{-1}W$ which is equivalent to $V^TV=1$.
The other elements of a discrete generation group can be constructed from
multiplications of $R$ and $T_n$. For example, for $n=2$
with $T\equiv T_2$, $T^2=1$ the list is $(R, T, RT, TR, RTR, TRT, RTRT=TRTR=-W)$.
(If one adds an element $I_1=diag(i,0,0)$ one can enforce the hierarchical
pattern where $b=a$.)

The maximal symmetry consistent with a mass matrix $M_B$ can be
generated from the $O(3)$ transformations encoded in $V$ (or
equivalently, $S$) and an abelian transformation $\nu\to e^{i\chi}\nu,
\ t\to e^{-2i\chi}t$. The subgroup of real transformations $S=S^*$ leaves
the matrix $M_{max}$ (eq. (\ref{A1})) invariant. This
corresponds to (abelian) rotations in the $(\nu_e,
\nu_A)$ plane. We conclude that maximal mixing with a degenerate mass
pattern can arise naturally from a large class of symmetries.

We next have to ask if such a generation symmetry is compatible
with a realistic mass pattern for the charged leptons. The leading
mass term has the form $l^{cT}d_l^*l$ where
$d_l$ denotes the doublet that contributes to
the charged lepton mass matrix $M_l$ in leading order.
(Typically, $d_l$ is not the leading $SU(2)_L$-breaking
doublet since $m_\tau\ll m_t$.) In general $l^{cT}=
(e^c,\mu^c,\tau^c)$ may transform under $S$ differently from $l^T=
(e,\mu,\tau)$, i.e. $l\to Sl,\ l^c\to S^{(c)}l^c$.  The doublet
transformation property $d_l^*\to(S^{(c)T})^{-1}d^*_lS^{-1}$
implies that possible
natural mass patterns depend crucially on the representation of $l^c$.
We concentrate here on the case of an $SO(3)$ generation group
where $l^c$ and $l$ belong both to triplet representations, i.e.
$S^{(c)}=S$. It is then convenient to work in the
canonical $SO(3)_G$ basis (\ref{9}) $\tilde
\nu=U^\dagger\nu,\tilde l=U^\dagger l, \tilde l^c=U^\dagger l^c$ where
the leading neutrino mass matrix is diagonal $\tilde M_B=U^TM_BU=m\eins$
and $SO(3)_G$ is represented by standard real orthogonal matrices $V$. A
leading doublet which only contributes to the $\tau$-mass yields in this
basis $\tilde M_l=m_\tau Y$ with $Y=U^Tdiag(0,0,1)U$ given by eq.
(\ref{A}).
The matrix $Y$ is traceless and symmetric and the leading
doublet should therefore belong to a 5 of $SO(3)_G$.
With respect to the $U(1)$-rotation $T(\varphi)$ (cf. eq. (\ref{10}))
\be\label{12}
\tilde T(\varphi)=U^\dagger T(\varphi)U=\left(\begin{array}{ccc}
1&0&0\\ 0&\cos \varphi&\sin \varphi\\ 0&
-\sin\varphi& \cos \varphi\end{array}\right)\ee
the matrix $Y$ transforms as
\be\label{13}
\tilde T(\varphi)^TY\tilde T(\varphi)=
e^{-2i\varphi}Y\ee
If the doublet transforms as $d_l^*\to e^{2i\varphi}d_l^*$, only
a lepton mass term $\tilde M_l\sim Y$ is allowed.

Let us denote the abelian charge related to the rotations $\tilde T
(\varphi)$ in the $(\nu_A,\nu_B)$ plane by $I_{3G}$. The first
generation $(\nu_e; e;e^c)$ has $I_{3G}=0$,
the second $(\nu_\mu;\mu;\mu^c)$ carries $I_{3G}=+1$ whereas
the third $(\nu_\tau;\tau;\tau^c)$ comes with  $I_{3G}=-1$. If
the $SU(2)_L$-doublet $d^*_l$ belongs to
the $I_{3G}=2$ component of an $SO(3)_G$-5-plet only the $\tau$ can
acquire a mass! On the other hand,
we assume that the leading $SU(2)_L$-triplet
belongs to an $SO(3)_G$-singlet with $I_{3G}=0$. The
qualitative difference between the hierarchical mass pattern
for the charged leptons and the degenerate pattern for the neutrinos
can simply be explained by the different transformation properties
of the leading doublet and triplet! We emphasize that the relative
angle of $\pi/4$ between the basis of eigenstates of $I_{3G}$
and the ``canonical $SO(3)_G$-basis'' arises very naturally
in this picture. It is the central ingredient for maximal
$\nu_\mu-\nu_\tau$-mixing.

For the maximal mixing scenario the different representations of the
doublet and triplet strongly disfavour the generation of
the leading contribution to the neutrino mass by the seesaw mechanism
\cite{seesaw}. In fact, the Majorana mass matrix for the
left-handed neutrinos can be written in the general form \cite{W1}
\be\label{14}
M_\nu=M^T_DM^{-1}_RM_D+M_L\ee
where $M_D$ is the Dirac mass matrix linking right- and left-handed
neutrinos, and $M_R$  is the mass matrix for the heavy
or ``right-handed''
neutrinos. Since $M_R$ transforms as an $SU(2)_L$-singlet, one
expects that its eigenvalues are much larger than the Fermi scale.
The seesaw mechanism is realized if $M_L$ can be neglected.
On the other hand, $M_L$ arises from the direct coupling of the
left-handed neutrinos to a scalar
$SU(2)_L$-triplet\footnote{In left-right symmetric unification
like $SO(10)$ the matrices
$M_L$ and $M_R$ are often proportional to each other.
This is the case whenever the $SU(2)_L$-triplet leading to
$M_L$ and the $SU(2)_L$-singlet leading to $M_R$ belong the the
same scalar multiplet.}.
As mentioned above, its expectation value
must be proportional to $M^2_W/M_t$ with $M_t$
a large mass scale corresponding to the mass of the scalar $SU(2)_L$
triplet \cite{Magg}. The degenerate neutrino mass pattern
can easily be realized if $M_L$ is the leading contribution to
$M_\nu$ with the $SU(2)_L$-triplet transforming as a singlet under
the $SO(3)_G$ generation group. In contrast, the Dirac mass term $M_D$
often has a similar structure as the charged lepton mass
matrix $M_l$ or the up-type quark mass matrix $M_u$ because it
also arises from the coupling to a doublet. As an example we
investigate the consequences of a leading-order behavior
$M_D\sim M_l\sim diag(0,0,1),\
M_R\sim M_L\sim W$. Since
$W^2=1$ one has $M^{-1}_R=\frac{1}{m_R}W$ and $M^T_DM^{-1}_RM_D=0$.
A Dirac mass contribution can therefore only arise at subleading
order. For $M_D=diag(g_e,g_\mu,g_\tau)\ ,\  g_e\ll g_\mu \ll g_\tau$ one
finds\footnote{Off-diagonal elements in $M_D$ can also
give an important contribution.}
\be\label{16}
M^T_DM_R^{-1}M_D=\left(\begin{array}{ccc}
g^2_e&0&0\\ 0&0&g_\mu g_\tau\\ 0&g_\mu g_\tau&0\end{array}\right)
m^{-1}_R\ee
This induces a mass split between $\nu_e$ on the one side
and the two linear combinations of $\nu_\mu$ and $\nu_\tau$ on
the other side. We conclude that $g_\mu g_\tau/m_R$ should be of
the order $\Delta m^2_s/m$ or smaller. The
dominant subleading correction to $M_\nu$ lifting the degeneracy
between $\nu_e$ and $\nu_A$ on one side and $\nu_B$ on the other side
should therefore arise from the expectation value of another
$SU(2)_L$-triplet contributing to $M_L$ or from a
correction\footnote{A matrix $M_R=m_RW+s_R\eins$, with real $m_R, s_R$ and
$|s_R|\ll|m_R|$ generates a Dirac mass contribution $\sim(g_\tau^2
s_R/m^2_R)Y$.
Since this contributes to $\Delta m^2_a$ and $\Delta m_s^2$
in comparable magnitude, the ratio $|s_R/m_R|$ should be small, e.g.
$s_R/m_R\approx g_\mu/g_\tau$, or $g_\tau^2/m_R$ must be much smaller
than $|b|$. For $g_\mu/g_\tau\approx m_c
/m_t\approx s_R/m_R\approx a/b$ one would find that the deviation
from maximal mixing is indeed small (cf. eq. (\ref{2}) with $c=a$) and
the neutrinos are almost degenerate in mass, with $m\approx
\frac{1}{2}\sqrt{b\Delta m_a^2/a}
\approx 0.35\ eV$. We observe that a typical value of $m=b=
M_W^2/M_t$ would imply $M_t\approx 2\cdot 10^{13}$ GeV whereas
$g_\mu g_\tau/m_R\approx \Delta m^2_s/m$
yields for $g_\tau\approx M_W, g_\mu/g_\tau\approx 10^{-2}$ a value
$m_R=4\cdot 10^{15}$ GeV. Smaller values of $g_\tau$ or $m$ lower
the value of $m_R$.} to
$M_R$. We have already discussed above that $M_L=bW+a\eins$ would
account for $\Delta m^2_a$ (cf. eq. (\ref{6})). This scenario is
realized if the next to leading $SU(2)_L$-triplet expectation value
remains invariant under the symmetries $R$ and $T$ or
the $(\nu_e,\nu_A)$ rotations generated by a
rotation $V$ in the 1-2 plane such that $S^TS=1,\ S^*=S$.

We emphasize, however, that the above discussion merely serves
as an example. If the unification group does not contain $SU(2)_R$,
there is no reason why the left-handed antineutrinos $\nu^c$ (which
are equivalent to the right-handed neutrinos) should be in the
same representation of the generation group as $l^c$. In consequence,
there is then no reason for a proportionality $M_D\sim M_l$. Similarly,
if the unification group does not contain $SU(4)_c$, there is no
relation between $M_D$ and the mass matrix $M_u$ for the
up-type quarks. An investigation of generation symmetries which are
compatible with $SU(5)$ reveals \cite{Bijnens} that $M_D$ often has a
generation structure which is quite different from $M_l$ or $M_u$. On
the other hand, there is also no reason why $M_D$ should be proportional
to $M_L$ and $M_R$. A generation of the maximal mixing
matrix $M_{max}$ by the seesaw-mechanism seems therefore unlikely.
Furthermore, a proportionality $M_L\sim M_R$ is not expected
if $\nu$ and $\nu^c$ belong to different representations of the
generation group.

We next sketch a possible mechanism\footnote{For a realistic
scenario of maximal $\nu_\mu-\nu_\tau$-mixing it is not
sufficient that the leading contributions to the mass matrices
have the corresponding structure. The pattern for the smaller
(nonleading) masses must remain compatible with maximal mixing.} for
small mass ratios $m_\mu/m_\tau$
and $m_e/m_\mu$. A mixing of nonleading doublets
coupling to muons and electrons with the leading doublet coupling
$\sim Y$ is only possible if all abelian subgroups of
$SO(3)_G\times U(1)_G$ with charges $\tilde q=I_{3G}+\alpha q$ are
spontaneously broken. Here $q$ denotes the charge corresponding
to the abelian generation group $U(1)_G$ with $q(\nu_e,\nu_\mu,\nu_\tau)
=q(e,\mu,\tau)=q_1,\ q(e^c,\mu^c,\tau^c)=q_2,q(d)=-(q_1+q_2)$.
The other doublets in the $5$ of $SO(3)_G$ have $I_{3G}=(-2,
-1,0,1)$ as compared to $I_{3G}=2$ for the  doublet $d^*_l$. This
difference in charge carries over to $\tilde q$
for arbitrary $\alpha$. An $SO(3)_G$ breaking
($SU(2)_L$-singlet) operator with fixed $\hat I_{3G}$ and nonzero
$q=Q$ leaves an $U(1)$ symmetry with $\alpha= -Q/\hat I_{3G}$
unbroken. Only the expectation value $\sim M_{\tilde q}$ of a second
such operator with $\tilde q=\tilde Q,\tilde Q\not=0$ can lead to a mixing
between the doublets with different $\tilde q$. The mixing of two
doublets whose difference in $\tilde q$ is given by $|\Delta\tilde q|
=p|\tilde Q|$ must be proportional $(M_{\tilde q})^p$. If $M
_{\tilde q}$ is smaller than a characteristic common heavy mass $M_d$
(the mass of the heavy doublets), this
induces a small parameter $\lambda=M_{\tilde q}/M_d$. For $\tilde Q
=\pm1$ a typical charged lepton mass pattern for our scenario is
\be\label{17}
M_l=m_\tau\left(\begin{array}{ccc}
\lambda^4&\lambda^3&\lambda^2\\ \lambda^3&\lambda^2&\lambda\\
\lambda^2&\lambda& 1
\end{array}
\right)\ee
where only the order of magnitude is indicated.
Similar patterns can arise for up- and down-type quark mass matrices.
This is, however, not the topic of this short note.
We also repeat here that $l^c$ and $\nu$ need not have the same
transformation properties with respect to the generation group.
(This would only be required for $SO(10)$ unification.) The relation
between symmetry and the generation pattern for the charged leptons may
be very different from the example discussed above. Nevertheless,
the basic observation that the $SU(2)_L$-doublets and -triplets
transform differently under the generation group is rather general.
This explains \cite{Bijnens} why the
generation structure of $M_L$ comes
often out quite different from $M_l,M_D$ or $M_u$.

Despite their different roots in generation symmetry representations
and breaking patterns the leading contribution to $M_\nu$ and the charged
fermion mass matrices show one striking similarity: The two smallest
mass eigenvalues are almost degenerate when compared with the scale
of the largest mass. This corresponds to an effective invariance
of the mass matrix under $U(1)$-rotations in the plane of the
two light eigenstates\footnote{These rotations are not necessarily
part of a genuine generation symmetry. It may happen that discrete
subgroups of the effective rotations belong to the generation
symmetry.}. Whereas for the charged fermions the effective rotations are
in the plane of the first two generations, the one in the
neutrino sector is in the $\nu_e-\nu_A$-plane which has an
angle of $\pi/4$ with respect to the $\nu_e-\nu_\mu$-plane. One
may call this difference in the planes of effective rotations the
``neutrino mismatch''. The question if such a neutrino mismatch
can arise naturally is crucial for the hierarchical neutrino mass
pattern (A) as well as for the next to leading order
contributions in the degenerate pattern (B). For the latter it
amounts to the question if a matrix of the type $M_{max}$ (eq. (\ref{A1}))
can be protected by symmetry. We have already identified a candidate
for such a symmetry, namely the discrete transformations $R:\nu_e
\leftrightarrow \frac{1}{\sqrt2}(\nu_\mu+\nu_\tau)$
and $T:\nu_\mu\to -\nu_\mu,\ \nu_\tau\to
-\nu_\tau$ (see eq. (\ref{10})).

We still have to ask if it is natural that the leading
$SU(2)_L$-triplets (there may be more than one, typically two
if $|a|\ll|b|\ $) are invariant under the transformations $R$ and $T$.
This has to be realized in a natural way in presence of the
singlets which induce the doublet mixing responsible for the masses
of the first two generations of charged fermions. (These singlets
have a fixed value of $I_{3G}$ in our example.) $R$ and $T$ invariance
for the induced triplet
is obviously realized for $SO(3)_G$-singlets. It is less trivial
for the (perhaps subleading) field $a$ responsible for
$\Delta m^2_a\not=0$. Typically the leading contributions
to the $SU(2)_L$-triplet expectation value $<t>$ is induced by
a linear term \cite{Magg} $\sim d_td_tt(s)$. Here $d_t$ is the
expectation value of the leading $SU(2)_L$-breaking doublet which
gives a mass to the top quark. The expectation value of an $SU(2)_L$-singlet
$s$ is needed if $d_td_tt$ is not invariant under the unification and
generation symmetries. In particular, if the unification symmetry includes
$B-L$ symmetry, the singlet $s$ has $B-L=-2$. A simple example
for a natural neutrino mismatch is a scenario where $d_t$ belongs to
a singlet\footnote{It is interesting to note in this context that the
first higher dimensional unification model with realistic fermion
charges, namely the six-dimensional $SO(12)$-model \cite{UNIF}, exhibits
many of the features discussed here if the ground state corresponds
to a particular monopole compactification with $SU(5)$ symmetry
$(n=3, m=p=1)$. The generation group $SO(3)_G\times U(1)_G$ has to be
broken in the vicinity of the compactification scale.
The $SU(2)_L$-doublet $(\nu,l)$ transforms as a
triplet under $SO(3)_G$
whereas $(t,b), t^c$ and $d_t$ are singlets.}
of the $SO(3)_G$-generation symmetry, whereas $s$ breaks
$SO(3)_G$, but preserves the discrete subgroup generated by
$R$ and $T$.

The pattern of neutrino mass eigenvalues depends on
details of the symmetry-breaking pattern of the generation symmetry.
In particular, all three neutrino masses are approximately degenerate
if the scale of $R$- or $T$-breaking is lower than the breaking of
$SO(3)_G$ (or a subgroup enforcing $a=0$). If this is not realized,
a similar size of $a$ and $b$ may lead to a pattern with the mass
of two of the neutrinos $(\nu_e,\nu_A)$ almost degenerate and
of the same order but not
approximately equal as the mass of the third neutrino $\nu_B$.

Finally, we address the question of the deviation from
maximal $\nu_\mu-\nu_\tau$-mixing or the size of $\epsilon=1-\sin^22\vartheta$.
We assume that the mixing between the second and third generation
in the charged lepton mass matrix is small, similar to the
quark mass matrices. A value $\vartheta_{23}\approx0.04$
contributes $\Delta\epsilon_l=4\vartheta^2_{23}\approx 6\cdot
10^{-3}$. In the neutrino sector, a small deviation from
$M_{max}$ (\ref{A1}) can be generated from the seesaw mechanism
$\sim$ $M^T_DM_R^{-1}M_D$. The terms responsible for a deviation
from maximal $\nu_\mu-\nu_\tau$-mixing are off-diagonal
in the standard $SO(3)_G$-basis where $\tilde M_{max}=U^TM_{max}U=
diag(b+a,b+a,b-a)$. They typically also contribute to
$\Delta m^2_s$. If the leading correction comes from the
element $(\tilde M_\nu)_{23}$ (and similar for $(\tilde M_\nu)_{13}$),
one finds $\Delta m^2_s=\Delta m^2_a\Delta\vartheta^2$
or $\Delta\epsilon_\nu=
4\Delta m^2_s/\Delta m^2_a\approx 4\cdot 10^{-3}$.
On the other hand, a leading correction in
$(\tilde M_\nu)_{12}$ would lead to maximal $\nu_e-\nu_A$-mixing.
Although this case would be
interesting in its own right, we concentrate here on
a small mixing of the electron neutrino and discard this
possibility. Combining $\epsilon=
(\sqrt{\Delta\epsilon_l}\pm\sqrt{\Delta\epsilon_\nu})^2$ we conclude
that the typical deviation from maximal mixing is below 1\%!.

It is not our aim here to device one particular realistic model
for a generation symmetry. We rather want to draw some general
conclusions from the outcome of this investigation:

(1) Maximal $\nu_\mu-\nu_\tau$-mixing can follow naturally
from a suitable generation symmetry. It is compatible with a
small mass squared difference $|m^2_{\nu_A}-m^2_{\nu_e}|
\approx 5\cdot10^{-6}eV^2$. If such a maximal mixing pattern is
found experimentally, this would give a strong hint for a nonabelian
generation symmetry. A minimal version of such a symmetry can be
generated from discrete transformations $R: \nu_e\leftrightarrow
\nu_A=\frac{1}{\sqrt2}(\nu_\mu+\nu_\tau)$ and
$T:\nu_{\mu,\tau}\to-\nu_{\mu,\tau}$.

(2) Small parameters appear in the deviation of $\sin^22\vartheta$
from one. In the neutrino sector the small quantity
is given by $M_{RT}/M$
where $M_{RT}$ is a typical scale
characterizing the spontaneous  breaking of the generation symmetry
(i.e. $R$ or $T$). Typically, this parameter also enters in the mass
split between $\nu_e$ and $\nu_A$. Discarding a large
contribution from the charged lepton mass matrix, the
deviation from maximal
$\nu_\mu-\nu_\tau$-mixing is expected to be tiny,
$1-\sin^22\vartheta\stackrel {\scriptstyle<}{\sim}
0.01$. This can be viewed
as a prediction of natural maximal mixing patterns and clearly
distinguishes them from scenarios where the mixing is large but
$1-\sin^22\vartheta$ is not related to a small symmetry-breaking
scale.

(3) In case of maximal $\nu_\mu-\nu_\tau$-mixing the leading
contribution to the mass matrix for the light neutrinos is
presumably due to the expectation value of a heavy $SU(2)_L$-triplet
scalar field rather than to the seesaw-mechanism.

(4) The three neutrino masses can be almost
degenerate if the generation
group is $SO(3)_G$ with neutrinos belonging to a
triplet whereas the leading $SU(2)_L$-triplet scalar
field transforms as a singlet. Maximal $\nu_\mu-\nu_\tau$-mixing
is then explained by the lepton mass matrix not being diagonal in the
standard $SO(3)_G$-basis.

In view of the important implications for an understanding
of possible generation symmetries stronger experimental limits
for the deviation of $\sin^22\vartheta$ from one would be of
great value.

\end{document}